# Deep learning within *a priori* temporal feature spaces for large-scale dynamic MR image reconstruction: Application to 5-D cardiac MR Multitasking

Yuhua Chen[1,2[0000-0002-4124-1148]], Jaime L. Shaw[2[0000-0003-2730-5982]], Yibin Xie[2[0000-0002-0333-567X]], Debiao Li[1,2[0000-0002-9334-8684]], and Anthony G. Christodoulou[2[0000-0002-9334-8684]]

[1] Department of Bioengineering, University of California, Los Angeles, CA 90095, USA
[2] Biomedical Imaging Research Institute, Cedars-Sinai Medical Center, CA 90048, USA
`chyuhua@ucla.edu`

**Abstract.** High spatiotemporal resolution dynamic magnetic resonance imaging (MRI) is a powerful clinical tool for imaging moving structures as well as to reveal and quantify other physical and physiological dynamics. The low speed of MRI necessitates acceleration methods such as deep learning reconstruction from under-sampled data. However, the massive size of many dynamic MRI problems prevents deep learning networks from directly exploiting global temporal relationships. In this work, we show that by applying deep neural networks inside *a priori* calculated temporal feature spaces, we enable deep learning reconstruction with global temporal modeling even for image sequences with >40,000 frames. One proposed variation of our approach using dilated multi-level Densely Connected Network (mDCN) speeds up feature space coordinate calculation by 3000x compared to conventional iterative methods, from 20 minutes to 0.39 seconds. Thus, the combination of low-rank tensor and deep learning models not only makes large-scale dynamic MRI feasible but also practical for routine clinical application[1].

## 1 Introduction

Dynamic imaging plays an important role in many clinical MRI exams, assessing tissue health by visualizing and/or measuring any of several dynamic processes within the body: cardiac motion, respiration, T1 or T2 relaxation, contrast agent dynamics, and more. However, dynamic MRI is a relatively slow imaging modality, necessitating acceleration methods which can reconstruct images from "incomplete" image data. Thanks to the demonstrated ability of multilayer neural networks to learn highly efficient image representations and to rapidly decode imaging data, deep learning has become a popular approach for reconstructing static images (or individual frames of dynamic images) from snapshots of incomplete imaging data. This time-independent approach to image reconstruction ignores relationships between different frames of the image sequence; as a result, dynamic deep learning methods have recently been introduced, sharing data across sequences between 30 and 200 frames long [1-3].

---

[1] This work was supported by NIH 1R01EB028146

A major unaddressed problem in dynamic deep learning MR image reconstruction is how to handle even longer image sequences—a challenge raised by two current trends in dynamic MRI. First, accelerated dynamic MRI techniques have pushed toward higher frame rates as a means of investigating dynamic processes with increased temporal resolution. Second, multi-dynamic/extra-dimensional MRI techniques [4-6] can now simultaneously image multiple dynamic processes at once by defining multiple "time dimensions" (i.e., time-varying independent variables), leading to exponential growth of frames per image sequence. These trends have led to massive problem sizes; for example, [6] reports image sequences between 20,000 frames and 140,000 frames long. Dynamic deep learning MR image reconstruction has not yet been demonstrated for image sequences of this size, largely due to GPU memory limitations and overfitting risks that stem from the increased numbers of weights required to exploit global temporal relationships by connecting all frames (even indirectly).

One memory-efficient approach for non–deep learning methods has been linear subspace modeling [7], a variant of low-rank imaging. There, rather than focusing on image sequences in the time-domain, images are modeled, reconstructed, and stored in a low-dimensional, subject-specific feature space. However, employing linear modeling alone is known to be less effective than combining this approach with sparse recovery inside the feature space [8, 9] (substantially slowing the reconstruction process) or replacing the linear subspace model with nonlinear manifold modeling [10] (discarding the memory benefits). These approaches all require slow iterative reconstruction, which compromises their practical application in the clinic, where reconstruction is expected to take seconds. This iterative reconstruction is especially slow for non-Cartesian acquisitions, which otherwise have excellent properties for dynamic imaging.

Here, we propose an approach combining the memory efficiency of linear subspace modeling with the quality of nonlinear manifold modeling. Our approach uses deep learning to recover image coordinates/feature maps within linear subspaces rather than directly recovering image sequences, allowing non-local temporal modeling across entire image sequences with many frames. Due to strong connections between deep learning and nonlinear manifold modeling [11], we interpret this approach as using deep learning to recover feature maps on a nonlinear manifold within a subject-specific linear subspace. Although the feature space changes with each subject and is determined from concurrent subject-specific auxiliary data, the transform learned by the network is generalized and can be applied in any linear subspace of the same dimensionality.

We evaluate this approach for 5-D cardiac MR Multitasking, a low-rank tensor approach with three time-dimensions (cardiac phase, respiratory phase, and inversion time) and >40,000 frames per image sequence, enabling non-ECG, free-breathing, myocardial T1 mapping [12]. For this evaluation, we compare mDCN [13] and the state-of-the-art DenseUnet [14] as example network architectures, training on 153 subjects. We further expanded the mDCN structure with dilated convolutional layers for a larger receptive field, which is proven to further improve the reconstruction performance. Our preliminary results show that our model can reduce image coordinate recovery time from 20 minutes for a single 2D slice into 0.39s, an >3000x speed improvement which puts online clinical deployment within reach. Both networks were capable of fast image reconstruction, but mDCN was faster, smaller, more accurate, and more precise than the more widely-used DenseUnet.

### 1.1 Contributions

- We developed a new deep learning approach to dynamic MR imaging reconstruction
- Specifically, we show that applying deep neural networks inside low-dimensional data-driven temporal feature spaces allows time- and memory-efficient learning of global temporal relationships, even for image sequences with >40,000 frames.
- We modified a highly efficient neural network mDCN to produce high-quality images with less time and memory usage than the more-popular DenseUnet.

## 2 Method

### 2.1 Background

Dynamic MRI produces a spatiotemporal image sequence $I(\mathbf{x}, \mathbf{t})$, a function of spatial location (denoted by vector $\mathbf{x} = [x_1, x_2, x_3]^T$ containing up to three spatial directions $x_i$) and one or more time dimensions (denoted by vector $\mathbf{t} = [t_1, t_2, \cdots, t_R]^T$ containing $R$ time-varying independent variables $t_i$). Here, we represent the discretized image sequence as a matrix $\mathbf{A} \in \mathbb{C}^{M \times N}$ with elements $A_{ij} = I(\mathbf{x}_i, \mathbf{t}_j)$, with $M$ spatial locations (voxels), and $N$ time points (frames). The matrix $\mathbf{A}$ is spatially encoded by the MR scanner, producing a vector of encoded data $\mathbf{d} = E(\mathbf{A})$, where $E(\cdot)$ typically comprises partial spatial Fourier encoding as well as additional spatial encoding from receiver sensitivity patterns.

The goal of image reconstruction is to recover the original image sequence $\mathbf{A}$ from the measured data $\mathbf{d}$, i.e., to find some operation $f$ such that $\mathbf{A} = f(\mathbf{d})$. Typically, it is not possible to sample $\mathbf{d}$ at or above the spatiotemporal Nyquist rate, so the data are undersampled. This leads to an ill-posed inverse problem, such that a general $f(\cdot) = E^{-1}(\cdot)$ does not exist. However, due to strong relationships between different image frames of $\mathbf{A}$, dynamic MR images lie on low-dimensional manifolds [7, 10]; reconstruction methods can exploit these temporal relationships to find an $f$ such that $f(E(\mathbf{A})) \approx \mathbf{A}$ for these images.

### 2.2 Subspace formulation

When the image sequence contains many frames, $\mathbf{A}$ can be rather large, presenting substantial computational challenges for designing an $f$ which produces $\mathbf{A}$ by exploiting global temporal relationships across all image frames. One memory-efficient approach to recover $\mathbf{A}$ while exploiting global temporal relationships has been to use linear subspace modeling: $I(\mathbf{x}, \mathbf{t}) = \sum_{\ell=1}^{L} u_\ell(\mathbf{x}) \varphi_\ell(\mathbf{t})$. This model implies that the matrix $\mathbf{A}$ has a low rank $L < \min(M, N)$ and can be efficiently factored as $\mathbf{A} = \mathbf{U}\boldsymbol{\Phi}$, where $\mathbf{U} \in$ $\mathbb{C}^{M \times L}$ has elements $U_{i\ell} = u_\ell(\mathbf{x}_i)$ and $\boldsymbol{\Phi} \in \mathbb{C}^{L \times N}$ has elements $\Phi_{\ell j} = \varphi_\ell(\mathbf{t}_j)$. In this formulation, the columns of the spatial factor $\mathbf{U}$ are feature maps containing the coordinates for $I(\mathbf{x}, \mathbf{t})$ within a "temporal feature space" spanned by $\{\varphi_\ell(\mathbf{t}_j)\}_{\ell=1}^{L}$. Sampling can be designed to include high-speed auxiliary data as a subset of $\mathbf{d}$ (sometimes called subspace training data or navigator data), from which the temporal feature space can be directly extracted via principal component analysis (PCA). This predetermines $\boldsymbol{\Phi}$ and updates the problem formulation to

$$\mathbf{d} = E(\mathbf{U}\boldsymbol{\Phi}) = E_{\boldsymbol{\Phi}}(\mathbf{U}),$$

where the new goal of image reconstruction is to find an $f_{\boldsymbol{\Phi}}$ such that $f_{\boldsymbol{\Phi}}(\mathbf{d}) \approx \mathbf{U}$.

Linear subspace modeling alone is typically not enough to produce a high-quality $\mathbf{A}$ from a highly under-sampled $\mathbf{d}$, so this approach is often combined by sparse recovery methods such as compressed sensing to find a $\mathbf{U}$ which itself has a sparse representation $\Psi(\mathbf{U})$, e.g., by solving the nonlinear reconstruction problem

$$f_{\boldsymbol{\Phi}}(\mathbf{d}) = \arg\min_{\mathbf{U}} \|\mathbf{d} - E_{\boldsymbol{\Phi}}(\mathbf{U})\|_2^2 + \lambda \|\Psi(\mathbf{U})\|_1. \tag{1}$$

In practice, (1) is solved $\mathbf{U}$ by "backprojecting" $\mathbf{d}$ onto the feature space as $E_{\boldsymbol{\Phi}}^*(\mathbf{d})$ (where * denotes the adjoint) or a pre-conditioned $E_{\boldsymbol{\Phi},\text{pc}}^*(\mathbf{d})$ and performing nonlinear iterative reconstruction such as the alternating direction method of multipliers (ADMM) upon the result entirely within the feature space [9]. This process is very slow, especially for non-Cartesian sampling patterns, for which $E_{\boldsymbol{\Phi}}(\cdot)$ comprises several non-invertible, non-separable multidimensional non-uniform fast Fourier transforms (NUFFTs) instead of invertible, separable FFTs.

## 2.3 Deep learning formulation

We aim to replace this slow, iterative process with a much faster deep learning network which exploits global temporal relationships while retaining the memory benefits of the linear subspace formulation. We constrain all processing of the network to occur within the *L*-dimensional temporal feature space (rather than the *N*-dimensional full temporal space), by: 1) backprojecting $\mathbf{d}$ onto the feature space as $\mathbf{U}_0 = E_{\boldsymbol{\Phi},\text{pc}}^*(\mathbf{d})$; then 2) passing $\mathbf{U}_0$ through a network to apply a learned reconstruction operator $g(\cdot)$:

$$\begin{aligned}\mathbf{U} &= f_{\boldsymbol{\Phi}}(\mathbf{d}) \\ &= g(\mathbf{U}_0) = g\bigl(E_{\boldsymbol{\Phi},\text{pc}}^*(\mathbf{d})\bigr).\end{aligned}$$

This approach is compatible with a wide range of network architectures for a wide range of dynamic imaging applications. The specific network structure and application we used to evaluate our approach will be described in the next section.

## 3 Experiment

### 3.1 MR Multitasking Application, Training Data, and Preprocessing

To demonstrate and evaluate our reconstruction approach, we designed a series of experiments to replace the MR Multitasking spatial factor estimation process, i.e., iterative reconstruction by Eq. (1). Multitasking for non-ECG, free-breathing quantitative T1 mapping of the heart produces a 5-D image $I(\mathbf{x}, \mathbf{t})$ with two spatial dimensions and three time dimensions: $\mathbf{t} = [c, r, \tau]^T$, where $c$ is cardiac phase, $r$ is respiratory phase, and $\tau$ is inversion time. There is an image frame for each combination of 344 T1-recovery time-points ($\tau$), 20 cardiac phases ($c$), and 6 respiratory phases ($r$), i.e., 344 x 20 x 6 = 41280 frames (a data size equivalent to 23 min of video at 30 fps).

In our experiments, we used non-Cartesian multichannel MRI, such that $E_{\Phi}(\mathbf{U}) = \Omega([\mathbf{SF}_{NU}\mathbf{U}]\Phi)$, where $\mathbf{F}_{NU}$ is the NUFFT, $\mathbf{S}$ applies coil sensitivity patterns, and $\Omega$ is the undersampling operator. We produced our network input according to $\mathbf{U}_0 = E^*_{\Phi,pc}(\mathbf{d}) = \mathbf{S}^{\dagger}\mathbf{F}^H_{NU}\mathbf{W}\Omega^*(\mathbf{d})\Phi^H$, where $\Omega^*(\cdot)\Phi^H$ transforms the data into the temporal feature space, $\mathbf{F}^H_{NU}\mathbf{W}$ regrids the non-Cartesian data by applying a density compensation function (the diagonal matrix $\mathbf{W}$) followed by the adjoint NUFFT $\mathbf{F}^H_{NU}$ (a process similar to filtered backprojection), and where the pseudoinverse $\mathbf{S}^{\dagger}$ performs a complex coil combination.

A total of 191 subjects' worth of raw dynamic MRI data were collected via on-site Siemens 3T MRI scanners, and split into 8:1:1 ratio as training, validation, and testing set. We used a rank of $L$=32 with image matrix size of 160 x 160, so the network was tasked with recovering a spatial factor $\mathbf{U}$ composed of 32 complex-valued 160 x 160 feature maps. In order to handle complex numbers, the real and imaginary parts of $\mathbf{U}$ were concatenated into a set of 64 real-valued 160 x 160 feature maps.

### 3.2 Training Parameters and Experiment Setting

The models were implemented in Tensorflow on a workstation with Nvidia GTX 1080TI GPU. Before feeding into the network, instance-wise normalization was done on both the input and label data by subtracting their mean and dividing by their standard deviation (std). We used Adam optimizer with a default learning rate of 1e-4 to minimize the L1 loss between the network output and the label data. Two network backbones were implemented and evaluated, one based on the mDCSRN [13] and the other based on the DenseUnet [14]. Both networks had the same densely-connected block setting (4 convolutional layers, 128 of growth rate, and ELU [15] nonlinear activation), whose details are shown in **Fig 1**. To evaluate the effects of regularization, we applied different L1 and L2 regularization on the weights to avoid overfitting. We then further investigated how the dilation rate affects the reconstruction quality. For all the experiment, models were trained for 300k steps, after which no further improvement was observed. The validation loss was monitored and the checkpoint with best validation loss was used to test the model performance.

To quantitatively analyze the results, we used three different measurements to compare results from the deep learning networks and the reference conventional iterative algorithm: 1) the normalized root mean square error (NRMSE) of the spatial factors; 2) three image similarity metrics, NRMSE, PSNR, and SSIM of the reconstructed image sequences for the whole cardiac cycle (20 frames) at the end-expiration (EE) respiratory phase, for inversion times corresponding to bright-blood and dark-blood contrast weighting (i.e., the two most clinically important qualitative image weightings); and 3) the accuracy and precision of T1 maps (i.e., the quantitative map produced by Multi-tasking). The timing of the network runtime was also recorded.

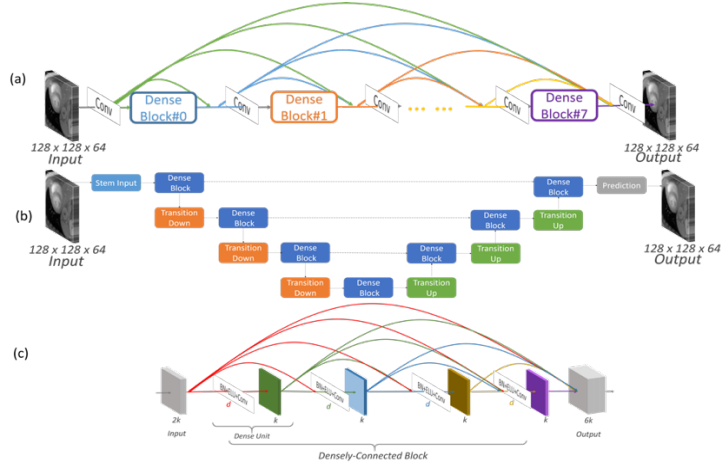

**Fig. 1.** Architecture of (a) the mDCN Network and (b) DenseUnet. Both have the same (c) Dense-Block (DB) with four 3x3 CONV layers. Dilation rate in CONV layers within a DB can vary and are notated as d1-4-8-1. DenseUnet has 4 resolution levels; the dilation rate of all layers is 1.

### 3.3 Results

The quantitative results of NRMSE on the spatial factor and image similarity metrics on cardiac cycles from different networks and configurations are shown in **Table 1** and **Table 2**. Hyperparameter searching was only done on the *validation set* to avoid overfitting to the *test set*. mDCN with dilation 1-4-8-1 and regularization scale of 0.01 outperformed all other networks. In general, mDCN outperformed the more popular DenseUnet in both image quality and speed. An example case of reconstructed MR images showing multiple contrasts and time dimensions is also shown in **Fig 2**.

|  | mDCN *d1-4-8-1* | | | DenseUnet | | |
|---|---|---|---|---|---|---|
| **L1&L2 Reg. Scale** | **No-Reg** | **1e-3** | **1e-2** | **No-Reg** | **1e-3** | **1e-2** |
| **Image Basis NRMSE** | 0.4460 (0.036) | 0.4324 (0.036) | **0.4302 (0.038)** | 0.4416 (0.035) | 0.4440 (0.033) | *0.4415 (0.034)* |
| **Cardiac Cycle SSIM** | 0.8329 (0.069) / 0.9164 (0.027) | **0.8524 (0.069)** / 0.9392 (0.026) | 0.8519 (0.070) / **0.9398 (0.037)** | 0.8450 (0.063) / 0.9292 (0.025) | 0.8252 (0.059) / 0.9033 (0.026) | *0.8368 (0.062) / 0.9128 (0.039)* |
| **Cardiac Cycle PSNR** | 29.44 (2.733) / 31.44 (3.052) | 30.51 (2.912) / 33.42 (3.081) | **30.70 (2.850)** / **33.74 (2.989)** | 29.97 (2.342) / 32.13 (2.243) | 29.16 (2.326) / 30.87 (2.385) | *29.70 (2.024) / 31.97 (2.021)* |
| **Cardiac Cycle NRMSE** | 0.1754 (0.057) / 0.1114 (0.044) | **0.1580 (0.056)** / 0.0894 (0.035) | 0.1554 (0.061) / **0.0863 (0.037)** | 0.1643 (0.052) / 0.1007 (0.030) | 0.1773 (0.048) / 0.1147 (0.028) | *0.1673 (0.050) / 0.1013 (0.025)* |
| **Runtime per case** | 0.39 s | | | *0.46s* | | |

**Table. 1.** The similarity metrics compared to iterative reconstruction for different L1 & L2 regularization scale on the *validation set*. The 0.01 scale gave the best performance.

|  | *Validation Set* | | | | *Test Set* | |
| --- | --- | --- | --- | --- | --- | --- |
|  | mDCN | | | DenseUnet | mDCN | DenseUnet |
| **Dilation** | **No Dilation** | **d1241** | **d1481** | **No Dilation** | **d1481** | **No Dilation** |
| **Image Basis NRMSE** | *0.4402(0.042)* | 0.4350(0.041) | **0.4302(0.038)** | 0.4415(0.034) | **0.4450(0.055)** | 0.4493(0.052) |
| **Cardiac Cycle SSIM** | *0.8474(0.065)* | 0.8472(0.071) | **0.8519(0.070)** | 0.8368(0.062) | **0.8619(0.070)** | 0.8381(0.079) |
|  | *0.9320(0.024)* | 0.9351(0.029) | **0.9398(0.037)** | 0.9128(0.039) | **0.9382(0.033)** | 0.9068(0.045) |
| **Cardiac Cycle PSNR** | *29.54(2.576)* | 30.07(3.126) | **30.70(2.850)** | 29.70(2.024) | **30.07(3.836)** | 29.20(3.374) |
|  | *31.20(3.382)* | 32.49(3.705) | **33.74(2.989)** | 31.97(2.021) | **32.07(4.300)** | 31.27(4.298) |
| **Cardiac Cycle NRMSE** | *0.1723(0.055)* | 0.1680(0.068) | **0.1554(0.061)** | 0.1673(0.050) | **0.1556(0.056)** | 0.1687(0.051) |
|  | *0.1161(0.047)* | 0.1050(0.057) | **0.0863(0.037)** | 0.1013(0.025) | **0.1052(0.053)** | 0.1127(0.043) |

**Table. 2.** Comparison between different dilation rates of mDCN on validation set. mDCN d1-4-8-1 with the largest effective receptive field gave the best scores on both *validation* & *test* set.

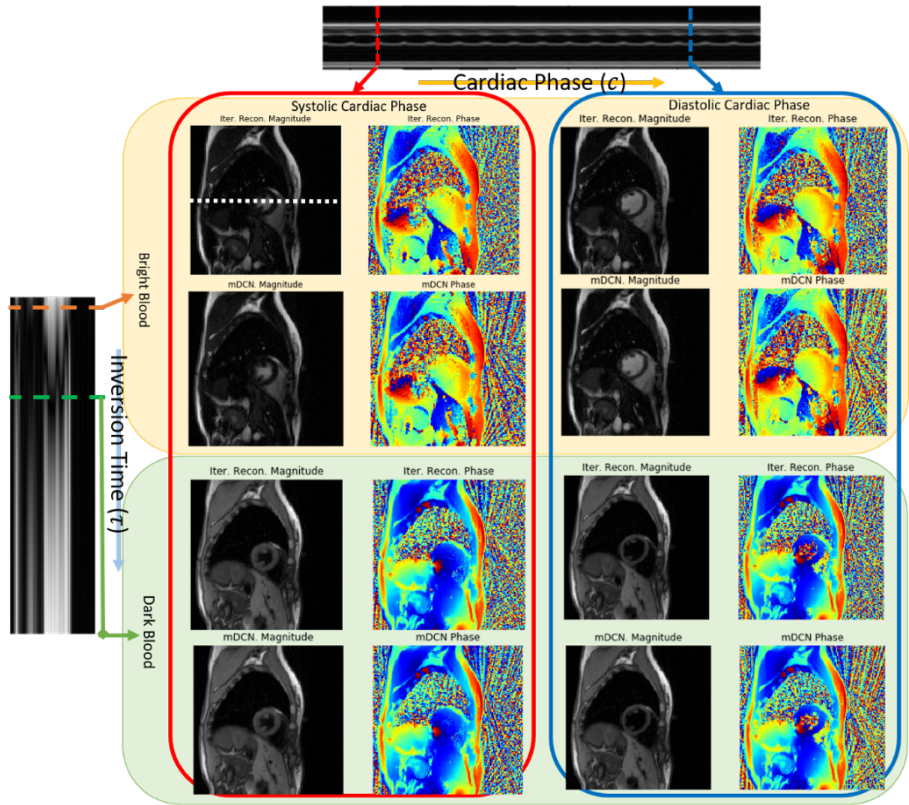

**Fig. 2.** Image samples from iterative reconstruction and mDCN in two dynamic phases: Inversion time ($\tau$), showing T1 recovery, and cardiac phase ($c$), showing cardiac motion.

Bland-Altman plots of the T1 fitting results and sample T1 maps are given in **Fig 3**. The mDCN maps were more accurate (smaller bias) and more precise (tighter limits of agreement) than DenseUnet. Neither network showed a statistically significant bias (mDCN: *p*=0.98; DenseUnet: *p*=0.36). The mDCN results were also more highly correlated to conventional results (mDCN: *R*=0.95, DenseUnet *R*=0.90).

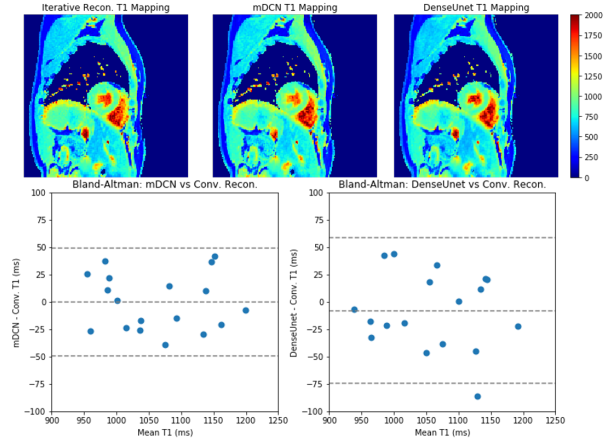

**Fig. 3.** T1 mapping results from conventional iterative reconstruction, mDCN, and DenseUnet. Bland-Altman analyses show that the mDCN is more accurate and precise than DenseUnet. In both cases, the limits of agreement are smaller than the ΔT1 of many diseases; e.g., median ΔT1 at 3 T between infarcted myocardium and remote myocardium in ST-segment elevation myocardial infarction (STEMI) patients was recently found to be 271 ms [16].

## 4    Conclusion

In this work, we have presented a fast, accurate approach to large-scale dynamic MRI reconstruction with global temporal modeling using a combination of low-rank modeling and deep learning. Our dilated mDCN provided similar-quality reconstructed images and T1 maps to conventional iterative methods, while reducing the time by >3000x. This reduction makes online reconstruction feasible for clinical application.

## References

1. Schlemper, J., Caballero, J., Hajnal, J.V., Price, A.N., Rueckert, D.: A Deep Cascade of Convolutional Neural Networks for Dynamic MR Image Reconstruction. IEEE Trans Med Imaging 37, 491-503 (2018)
2. Qin, C., Schlemper, J., Caballero, J., Price, A.N., Hajnal, J.V., Rueckert, D.: Convolutional Recurrent Neural Networks for Dynamic MR Image Reconstruction. IEEE Trans Med Imaging 38, 280-290 (2019)
3. Biswas, S., Aggarwal, H.K., Jacob, M.: Dynamic MRI using model-based deep learning and SToRM priors: MoDL-SToRM. Magn Reson Med (2019)
4. Feng, L., Axel, L., Chandarana, H., Block, K.T., Sodickson, D.K., Otazo, R.: XD-GRASP: Golden-angle radial MRI with reconstruction of extra motion-state dimensions using compressed sensing. Magn Reson Med 75, 775-788 (2016)
5. Cheng, J.Y., Zhang, T., Alley, M.T., Uecker, M., Lustig, M., Pauly, J.M., Vasanawala, S.S.: Comprehensive multi-dimensional MRI for the simultaneous assessment of cardiopulmonary anatomy and physiology. Sci Rep 7, 5330 (2017)

6. Christodoulou, A.G., Shaw, J.L., Nguyen, C., Yang, Q., Xie, Y., Wang, N., Li, D.: Magnetic resonance multitasking for motion-resolved quantitative cardiovascular imaging. Nature Biomed Eng 2, 215-226 (2018)

7. Liang, Z.-P.: Spatiotemporal imaging with partially separable functions. Proc IEEE Int Symp Biomed Imaging, pp. 988-991 (2007)

8. Lingala, S.G., Hu, Y., DiBella, E., Jacob, M.: Accelerated dynamic MRI exploiting sparsity and low-rank structure: k-t SLR. IEEE Trans Med Imaging 30, 1042-1054 (2011)

9. Zhao, B., Haldar, J.P., Christodoulou, A.G., Liang, Z.-P.: Image reconstruction from highly undersampled (k,t)-space data with joint partial separability and sparsity constraints. IEEE Trans Med Imaging 31, 1809-1820 (2012)

10. Poddar, S., Jacob, M.: Dynamic MRI Using SmooThness Regularization on Manifolds (SToRM). IEEE Trans Med Imaging 35, 1106-1115 (2016)

11. Bengio, Y., Courville, A., Vincent, P.: Representation learning: A review and new perspectives. IEEE Trans Pattern Anal Mach Intell 35, 1798-1828 (2013)

12. Shaw, J.L., Yang, Q., Zhou, Z., Deng, Z., Nguyen, C., Li, D., Christodoulou, A.G.: Free-breathing, non-ECG, continuous myocardial $T_1$ mapping with cardiovascular magnetic resonance Multitasking. Magn Reson Med 81, 2450-2463 (2019)

13. Chen, Y., Shi, F., Christodoulou, A.G., Xie, Y., Zhou, Z., Li, D.: Efficient and accurate MRI super-resolution using a generative adversarial network and 3D multi-Level densely connected network. In: MICCAI, pp. 91-99. Springer, (2017)

14. Jégou, S., Drozdzal, M., Vazquez, D., Romero, A., Bengio, Y.: The one hundred layers tiramisu: Fully convolutional densenets for semantic segmentation. In: Proceedings of the IEEE Conference on Computer Vision and Pattern Recognition Workshops, pp. 11-19. (2017)

15. Clevert, D.-A., Unterthiner, T., Hochreiter, S.: Fast and accurate deep network learning by exponential linear units (elus). arXiv preprint arXiv:1511.07289 (2015)

16. Kali, Avinash, Eui-Young Choi, Behzad Sharif, Young Jin Kim, Xiaoming Bi, Bruce Spottiswoode, Ivan Cokic et al. "Native T1 mapping by 3-T CMR imaging for characterization of chronic myocardial infarctions." JACC: Cardiovascular Imaging 8, no. 9 1019-1030 (2015).